# Universal superconducting precursor in perovskite-based oxides

D. Pelc[1,*], Z. Anderson[1], B. Yu[1], C. Leighton[2], and M. Greven[1,*]

[1] School of Physics and Astronomy, University of Minnesota, Minneapolis MN-55455, USA

[2] Department of Chemical Engineering and Materials Science, University of Minnesota, Minneapolis MN-55455, USA

* Correspondence to: dpelc@umn.edu, greven@umn.edu

**A pivotal challenge posed by unconventional superconductors is to unravel how superconductivity emerges upon cooling from the generally complex normal state. Some of the most prominent unconventional superconductors are oxides: strontium titanate, strontium ruthenate, and the cuprates exhibit greatly different superconducting transition temperatures $T_c$, and although their respective superconducting pairing mechanisms remain unknown, they are thought to differ as well[1-5]. Here we use nonlinear magnetic response, a probe that is uniquely sensitive to the superconducting precursor, to uncover remarkable universal behavior in these three distinct classes of superconductors. We find unusual exponential temperature dependence of the diamagnetic response above the transition temperature $T_c$, with a characteristic temperature scale that strongly varies with $T_c$. We correlate this scale with the sensitivity of $T_c$ to local stress, and show that it is influenced by intentionally-induced structural disorder. The universal behavior is therefore caused by intrinsic, self-organized structural inhomogeneity, inherent to the oxides' perovskite-based structure[6,7]. The prevalence of such inhomogeneity has far-reaching implications for the interpretation of electronic properties of perovskite-related oxides in general.**

We probe the superconducting precursor regime above $T_c$ by measuring nonlinear magnetic response. The magnetic response of a material to an external magnetic field beyond the linear approximation may be written as

$$M = \chi_1 H + \chi_3 H^3 + \cdots \quad (1)$$

where the $\chi_i$ are the susceptibility tensors, $M$ is the magnetization, and $H$ is the applied magnetic field. We assume that time-reversal symmetry is obeyed, and hence do not consider even-power terms. Superconducting transitions are clearly seen in linear response because of the diamagnetism associated with the superconducting state, but above $T_c$ the normal-state properties also contribute to $\chi_1$. The third-order response is virtually zero in the normal state, however, and thus uniquely sensitive to the superconducting precursor. Our experiments are performed with an alternating field, $H = H_0 \cos(\omega t)$, applied via an excitation coil, and the response is detected with a pair of pickup coils at frequencies $\omega$ (linear response) and $3\omega$ (nonlinear response). A typical excitation amplitude is $H_0 = 1$ Oe. Phase-sensitive detection enables frequency selectivity, with extremely high sensitivity in the kHz range (see Methods). The present experiment should be contrasted with previous radio-frequency measurements of cuprate nonlinear response, which were analysed in terms of conductivity due to the much larger induced electric fields[8]. We discuss all results here in terms of the raw third-harmonic signal, $M_3$.

Nonlinear response measurements were performed on the compounds $SrTiO_3$ (STO), $Sr_2RuO_4$ (SRO), $La_{2-x}Sr_xCuO_4$ (LSCO) and $HgBa_2CuO_{4+\delta}$ (Hg1201). These perovskite-based oxides are of tremendous scientific interest: Superconductivity in doped STO occurs at some of the lowest known charge carrier densities and, more than five decades after its discovery[9], there has been a recent upsurge of research activity[1,2,3]; SRO is a candidate for unconventional spin-triplet superconductivity[4]; and LSCO and Hg1201 are two of the most prominent cuprate superconductors[5]. In all cases, the superconducting pairing mechanism remains unknown. These oxides are quite different in many respects, e.g., their optimal transition temperatures span two orders of magnitude: $T_c \approx$ 400 mK (STO), 1.5 K (SRO), 38 K (LSCO) and 95 K (Hg1201). Remarkably, we find the same behaviour of the superconducting precursor in all four systems (Fig. 1). The data clearly show exponential temperature dependence that spans at least two orders of magnitude in $M_3$, with the temperature range limited only by detection noise. Furthermore, there is no appreciable change with excitation frequency. The characteristic scale $\Xi$ of the exponential decrease is obtained from fits to $M_3 \sim e^{-T/\Xi}$, and the resulting values are plotted in Fig. 2a. Clearly, $\Xi$ increases with $T_c$, but at a significantly sublinear rate. The values should be viewed with some caution, however, since for STO, LSCO and Hg1201, $T_c$ depends on the carrier-doping level. We work here with nearly optimally-doped crystals (with the highest $T_c$) as representative examples.

Regardless of interpretation, the simple universal behaviour is remarkable. First, the extended exponential tail is distinctly different from what is expected and observed in conventional superconductors. We have measured $M_3(T)$ for various conventional superconductors, and indeed find that the nonlinear response decays quickly below the noise level and exhibits power-law behavior consistent with Ginzburg-Landau theory and its extensions[10,11] (Fig. S1). Second, the oxides feature varied transition-metal ions, a wide range of optimal $T_c$ values, and pairing mechanisms that are thought to be quite different. STO superconductivity may be phonon[1]- or plasmon[2]-mediated, or related to ferroelectric criticality[3], SRO is a possible triplet superconductor with a potentially important role played by ferromagnetic fluctuations[4], and LSCO and Hg1201 are members of the cuprate high-$T_c$ family known to exhibit unusual pseudogap phenomena[5]. Moreover, the crystal (and electronic) structures differ as well. STO is a cubic perovskite that transforms to tetragonally-distorted on cooling, whereas SRO, LSCO and Hg1201 are in the perovskite-related, layered Ruddlesden-Popper series. However, all compounds feature a common perovskite unit with transition-metal-oxygen octahedra and the associated general tendency toward structural deformation[6,7]. Both STO and LSCO exhibit a well-defined symmetry-lowering structural transition on cooling[12,13], Hg1201 is nominally tetragonal, but with a sizable distribution of local crystal fields[14], whereas SRO is close to an octahedral rotation instability[15].

The observed exponential behaviour is highly unusual, and similar features are scarce in condensed matter physics. Exponential tails can appear in an ordered system with a coupling of the order parameter to local inhomogeneity; rare regions with exceptionally high local ordering temperatures then cause the tails[16]. A similar effect occurs in the dynamics of electrons in a random lattice potential, giving rise to exponential tails in the density of states of disordered semiconductors[17,18]. The same physical arguments can be adapted to superconductors, if the inhomogeneity occurs at length-scales comparable to the superconducting coherence length[19]. In the cuprates, the exponential temperature dependence of the precursor superconductivity was recently characterized using several complementary techniques – linear[20] and nonlinear[8] conductivity, as well as torque magnetization[21] – and it was found that a simple model based on spatial $T_c$ inhomogeneity agrees well with the data. By extension, this indicates that similar distributions universally exist in perovskite-based superconductors. Furthermore, since the only certain common feature shared by the distinct superconductors studied here is structure, the underlying origin of the electronic inhomogeneity must be related to structural inhomogeneity. We emphasize that this is at most indirectly related to point defects, since the different oxides have varying levels of defect concentrations (their concentration in SRO is minimal, although ruthenium inclusions can influence the results; see Supplementary Information). Rather, the inhomogeneity must be caused by intrinsic local stress accommodation that leads to atomic displacements. In metals and alloys, structural transitions that are displacive (i.e., non-diffusive) and symmetry-lowering are known as martensitic transitions[22]; they are accompanied by local deformations across multiple length scales, which can exist even if there is no macroscopic transition[7]. Analogous deformations in the perovskite-based oxides could be the underlying cause of the extended precursor regime that we observe.

We test this hypothesis in two ways. First, we emphasize the following simple observation. In all studied oxides, $T_c$ quite strongly depends on lattice deformation. A straightforward means to quantify this is the derivative of $T_c$ with respect to uniaxial stress, $dT_c/dP_i$, where $P_i$ is the pressure applied in a given crystallographic direction $i$. If lattice deformation at the nanoscale (i.e., at scales comparable to the superconducting coherence length) is the cause of inhomogeneity that gives rise to exponential behaviour, we would still expect $dT_c/dP_i$ to be a reasonable measure of the coupling between local $T_c$ and local (nanoscale) lattice strain. Therefore, the scale $\Xi$ should be related to $dT_c/dP_i$. Indeed, as shown in Fig. 2b, upon normalizing $\Xi$ by the values of $dT_c/dP_i$ for the four different oxides[23-25] we obtain a nearly universal value. $dT_c/dP_i$ is negative for STO, but since we only use it as a measure of coupling to lattice deformation, the absolute value is relevant for determining the $T_c$ inhomogeneity scale. We thus demonstrate that

$$\Xi \approx A \left| \frac{dT_c}{dP_i} \right|, \qquad (2)$$

where the constant $A \sim 100$ MPa plays the role of a characteristic stress scale, and may be interpreted as a measure of the intrinsic local stress distribution width. Since the elastic moduli of the oxides[26,27] are on the order of 100 GPa, the associated lattice deformations amount to 0.1% of the lattice parameter, or about 0.1 pm. This estimate may be crude, but it serves to show that the relevant deformations at the atomic level are, on average, very small. Yet they may be within reach of high-precision neutron scattering[28] and electron microscopy[29] techniques. Notably, the simple relation (2) is borne out in previous studies of the cuprates. It is known from thermodynamic studies[25] that $dT_c/dP_i$ is nearly universal in the cuprates, and a characteristic precursor scale similar to $\Xi$ was also found to be universal[8,20,21], in accordance with (2). Importantly, the oxides studied here have very different superconducting coherence lengths of ~ 75 nm (SRO[4]), ~ 30 nm (STO[30]) and 1-3 nm (cuprates[31]). Since $A$ appears to be universal, the relevant local strain should have a scale-free structure to cause the same effect on $T_c$.

A more direct test of the structural origin of the exponential tails is to study the influence of intentionally-introduced structural inhomogeneity. A LSCO single crystal was thus subjected at room temperature to in-plane uniaxial pressure beyond the elastic regime (Fig. 3a,b), reaching stresses comparable to $A$. A relatively low yield stress is observed, followed by plastic deformation and hysteresis; such features are not uncommon in compressively-deformed single crystal oxide ceramics[26], but to our knowledge have not been explored in the cuprates. The deformation must induce structural inhomogeneity (defects) to accommodate plasticity, beyond what is already present in the undeformed sample. We then compare nonlinear magnetization measurements on the same sample before and after applying 50 MPa pressure (Fig. 3c). The exponential slope changes dramatically, which demonstrates a clear relation to structural inhomogeneity. The ratio of the slopes in Fig. 3c is 1.7; if $\Xi$ is simply proportional to the local stress distribution width, this change is roughly what could be expected from the naïve addition of 50 MPa of induced stress to 100 MPa of pre-existing internal stress. Remarkably, the applied stress creates regions with locally-increased $T_c$ compared to the unstressed sample.

Our results have at least four far-reaching implications. First, we have uncovered an unanticipated common feature among perovskite-based superconductors, present regardless of the significant differences in their electronic properties. Second, the strong link between the characteristic temperature scale $\Xi$ and the coupling of $T_c$ to lattice strain suggests a common structural origin, rooted in the propensity for local deformation inherent to perovskites. Such effects have already been explored for some systems, especially the colossal magnetoresistive manganites[32], but should be relevant for perovskite-based materials in general, be they bulk superconductors, ferromagnets, or ferroelectrics, or films/heterostructures[33]. This insight is already beginning to be applied to the analysis of experiments on cuprates[8,20,21,34]. In the Supplementary Information, we demonstrate for SRO that the exponential temperature dependence is also apparent in a rather different observable: the electronic specific heat. Third, we show that it is possible to manipulate the inhomogeneity using mechanical stress, creating regions of locally increased $T_c$. Such experiments open up a new avenue for the study of unconventional superconductors; they should be invaluable in elucidating the nature of the intrinsic inhomogeneity with structural probes, and pave the way toward stress-engineering of superconducting properties. Finally, the observation of a universal intrinsic strain $A$ suggests that the underlying local deformations are similar in bulk perovskite-related materials, at least at the scale of the superconducting coherence length. Since for the oxides considered here these lengths differ by more than an order of magnitude, it is plausible that the deformations are self-organized and scale-free to some extent. It has been suggested that interstitial oxygen atoms in some cuprates form scale-invariant structures[35]. Our work indicates that such phenomena are a general property of perovskite-based materials, and it will be interesting to understand if they are pervasive in other unconventional superconductors.

**Methods**

*Samples.* The bulk STO sample is a single crystal cut from a commercially available wafer doped with 1 at. % of Nb, with the structure, part-per-million purity[36], and normal state characterized in previous work[37]. The bulk SRO crystals were grown by the traveling-solvent floating-zone technique and characterized by X-ray diffraction, showing a high degree of crystalline perfection. The high value of $T_c$ = 1.51 K indicates that the crystal in Fig. 1 is in the clean limit, i.e., that the level of point disorder is minimal[4]. The LSCO samples were grown by the traveling-solvent floating-zone method and characterized with X-ray diffraction. The sample for which the data in Fig. 1 were obtained was also characterized with energy-dispersive x-ray spectroscopy, where the doping level $x = 0.14$ was found to be homogeneous across the sample. All LSCO samples were annealed post-growth in air at 800°C for 24 hours. The Hg1201 sample was grown using an encapsulation method as described previously[38] and annealed in flowing oxygen gas for one month. The Pb and V samples are commercially-available 99.9% purity polycrystalline samples cut into rectangular bars, while the Nb sample is a commercially-available single crystal. The $Sn_{97}Ag_3$ sample is a standard soldering alloy.

*Nonlinear magnetization.* In order to measure the third-order nonlinear susceptibility, two different experimental setups were used. For measurements below 2 K, we employed a $^3$He recirculating refrigerator, while for measurements at higher temperatures we constructed a dedicated probe used within a Quantum Design, Inc., Magnetic Property Measurement System (MPMS). The $^3$He setup uses a set of one excitation and two detection coils, with the detection coils connected in such a way as to cancel the excitation field. The excitation current is supplied by a Stanford Research DS360 ultra-low distortion generator, and for detection we use an EG&G 5302 lock-in amplifier with transformer input (below 10 kHz) and with direct pre-amp input (above 10 kHz). The additional distortion from the transformer and pre-amp is negligible in the frequency range of interest. The sample and coils are shielded by a double Pb can superconducting shield, and we estimate the residual field in the sample space to be on the order of 0.01 Oe. The MPMS probe only contains the two matched detection coils, while for excitation we use the built-in coil for AC susceptometry. The MPMS is otherwise only used for temperature control. Both excitation and detection in the MPMS setup are provided by a Signal Recovery 7265 lock-in amplifier; we do not use a transformer. The sample is shielded by an external mu-metal shield, with the superconducting magnet of the MPMS quenched (reset) at low field before the experiments. The residual field in this configuration is estimated to be no larger than 0.1 Oe. The excitation fields in all experiments are on the order of 1 Oe (without corrections for demagnetization factors).

*Uniaxial stress.* Uniaxial pressure was applied to the sample at 300 K with a cell that uses helium gas pressure to create a force on a piston which, in turn, compresses the sample. The design enables independent control and measurement of applied pressure and sample strain. Tungsten carbide composite blocks are used to transfer the force to the sample, and the sample deformation is measured with a linear variable transformer (LVT) that consists of two concentric coils wound on ceramic formers attached to the piston and stationary end, respectively. The sample used in these experiments has a cross-section of 0.64 mm$^2$ and is 1.26 mm long; the cell force-He pressure ratio is 61 N/bar. Therefore, approximately 0.5 bar was needed to obtain the 50 MPa stress in the sample studied here.

**Acknowledgments**

We thank W. Zimmermann for the donation of the $^3$He fridge used (and for invaluable advice on its operation), A. S. Gibbs, C. W. Hicks and A. P. Mackenzie for providing the $Sr_2RuO_4$ samples, N. Bielinski for help in preparing the LSCO samples, S. Griffitt for assistance in constructing the susceptometer probes, A. Najev and M. Lukas for assistance with the uniaxial pressure measurements, and N. Barišić and B. Shklovskii for discussions. The work at the University of Minnesota was funded by the Department of Energy through the University of Minnesota Center for Quantum Materials under DE-SC-0016371.

**Author Contributions**

DP and MG initiated the work. DP designed the probes and performed the nonlinear response measurements, with assistance from ZA and BY. ZA and BY grew and characterized the Hg1201 and LSCO samples, respectively. CL provided and characterized the STO sample. DP carried out the data analysis. DP and MG wrote the manuscript, with input from all authors.

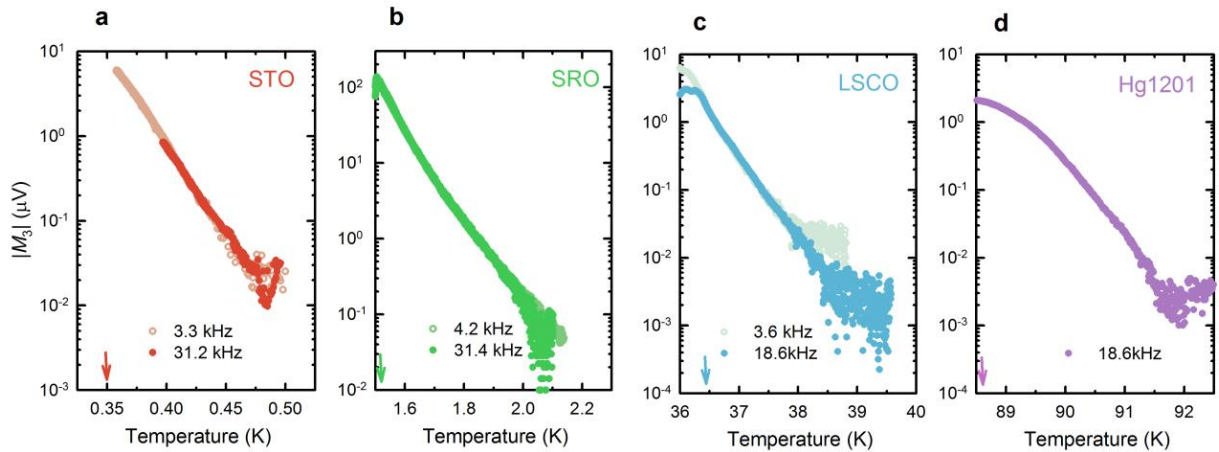

**Figure 1 | Nonlinear response of four perovskite-based superconductors above $T_c$. a**, Strontium titanate (STO) doped with 1 at. % of Nb, with a $T_c$ of 350 mK. **b**, Strontium ruthenate (SRO), with a $T_c$ of 1.51 K. **c**, Slightly underdoped lanthanum strontium cuprate (LSCO), with a $T_c$ of 36 K (Sr doping level of 14%). **d**, Slightly underdoped mercury barium cuprate (Hg1201), with a $T_c$ of 88.5 K. Arrows indicate the respective $T_c$ values. All four oxides show an exponential tail that stretches over at least two orders of magnitude, and with a slope that is approximately independent of excitation frequency. The low-frequency data are raw data, whereas the high-frequency data were multiplied by a constant to coincide with low-frequency measurements. The constants depend on the setup and sample penetration depth, and are 0.7, 5 and 0.2 for STO, SRO and LSCO, respectively (Hg1201 was only measured at one frequency).

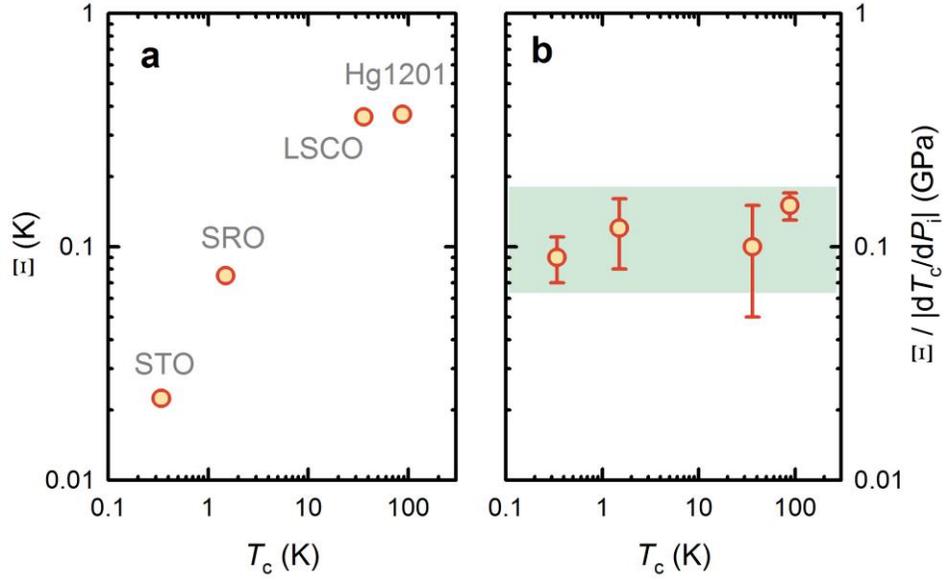

**Figure 2 | Characteristic pre-pairing scale. a,** The slopes $\Xi$ of the exponential dependences from Fig. 1, plotted against the $T_c$ values of the studied compounds. The errors are obtained from exponential fits and are smaller than the symbol size. **b,** $\Xi$ normalized to the derivative of $T_c$ with regard to uniaxial stress (Table 1). For STO, the derivatives $P_a$, $P_b$ and $P_c$ with respect to the three principal crystallographic directions are similar due to the nearly cubic structure, and for SRO, LSCO and Hg1201, we have taken values of $dT_c/dP_i$ with the pressure applied within the $RuO_2$ and $CuO_2$ planes. In SRO, the dependence of $T_c$ on $P_a$ is strongly nonlinear, and we have taken the average slope between $T_c = 1.5$ K and $T_c = 2$ K. The nonlinearity may be the cause of the slight curvature seen in Fig. 1b. The turquoise band indicates the estimated uncertainty range (see also Table 1).

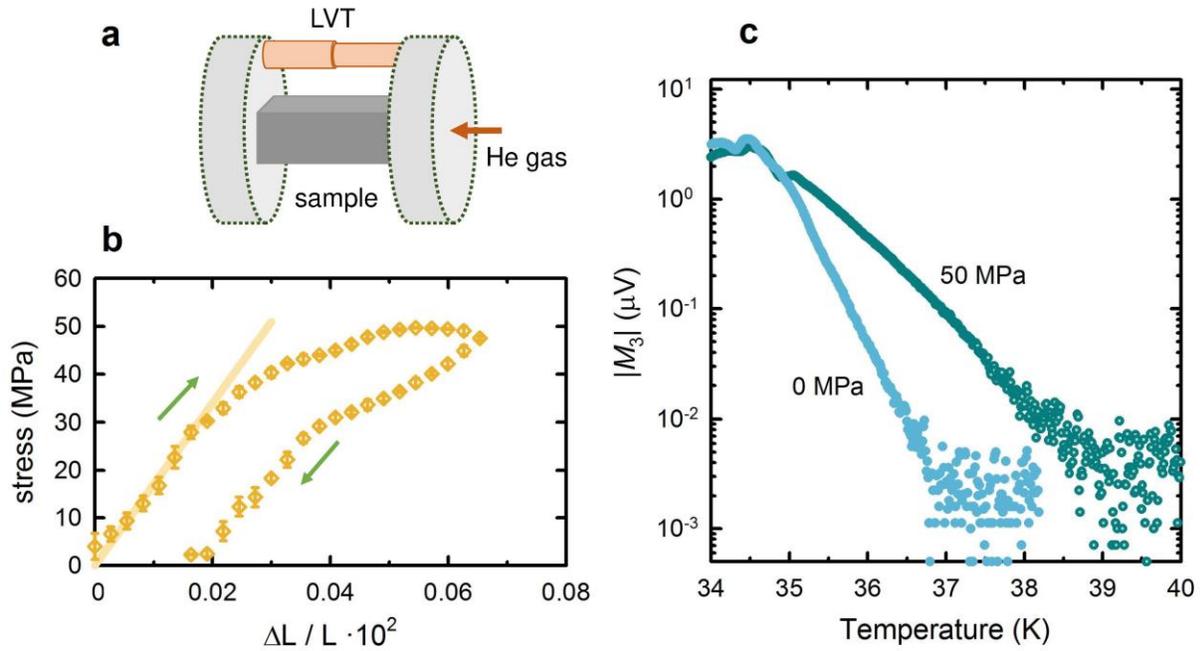

**Figure 3 | Influence of stress-induced inhomogeneity. a,** Schematic representation of the uniaxial pressure experiment, where a controlled force is applied to the sample with a helium gas piston (right disk), while the other side is static (left disk). The sample deformation is measured independently using a linear variable transformer (LVT). **b,** Room temperature stress-strain diagram of a LSCO sample with Sr doping level slightly below 14% (different growth batch than the sample in Fig. 2). The deviation from linear behaviour and the hysteresis clearly show that the plastic regime is reached, thus inducing structural inhomogeneity (defects) to accommodate plasticity. Multiple pressure readings were acquired at each deformation, and error bars are one standard deviation from the mean. The line corresponds to a Young modulus of 170 GPa, in good agreement with the value obtained from ultrasonic measurements of the elastic tensor components, 181 GPa (ref. 27). **c,** Nonlinear magnetization measurements before (full circles) and after (empty circles) the application of stress at room temperature. The stress-induced inhomogeneity has a strong influence on the slope of the exponential temperature dependence, establishing its structural origin. Note that the slope of the unstressed sample is the same as in Fig. 1, demonstrating the high level of repeatability between different growth batches.

| compound | Ξ (mK) (fit range) | $|dT_c/dP_i|$ (K/GPa) | coherence length (nm) |
|---|---|---|---|
| STO | 22.4 ± 0.1  (0.36 K - 0.47 K) | 0.25 ± 0.05 (ref. 23) | ~30 (ref. 30) |
| SRO | 75.0 ± 0.6  (1.55 K – 2 K) | 0.6 ± 0.2 (ref. 24) | ~75 (ref. 4) |
| LSCO | 358 ± 1    (36.5 K – 38.5 K) | 3.7 ± 1.3 (ref. 25) | ~2 (ref. 31) |
| Hg1201 | 367 ± 2    (89.5 K – 91.5 K) | 2.3 ± 0.3 (ref. 25) | ~2 (ref. 31) |

**Table 1 | Characteristic temperatures and uniaxial pressure derivatives for the oxide superconductors.** For SRO, LSCO and Hg1201, the average in-plane values of the derivatives are used; the large error margin for LSCO reflects the fact that the two principal in-plane values differ by a factor of 2. In SRO, the dependence of $T_c$ on $P_a$ is strongly nonlinear, and we have taken the average slope between $T_c$ = 1.5 K and $T_c$ = 2 K by simply fitting a linear function to data from two samples in ref. 24.

## Supplementary Information

*Conventional superconductors.* The study of superconducting fluctuations in conventional systems is hampered by the extremely small temperature range above $T_c$ where such effects can be observed in nonlinear response. In order to study nonlinear effects in conventional superconductors, we chose four different materials: V ($T_c$ = 5.3 K, type II), Nb ($T_c$ = 9.31 K, type II), Pb ($T_c$ = 7.2 K, type I) and a SnAg alloy ($T_c$ = 3.67 K, type I). V, Pb and SnAg were polycrystalline samples, while Nb was a single crystal (see also Methods). The measured third-order signals are shown in Fig. S1 on a relative temperature scale $\epsilon = (T - T_c)/T_c$, compared to SRO (from Fig. 1), which was chosen because of the very good signal-to-noise ratio. Clearly, the response in conventional superconductors quickly decreases above $T_c$, and displays a power-law tail over at least 1-2 orders of magnitude in $M_3$. Mean-field Ginzburg-Landau theory predicts that the third-order magnetization should decrease as $\epsilon^{-5/2}$ [11] for small applied magnetic fields and at relatively large $\epsilon$; this prediction is essentially borne out by these data. Nb is an exception, as it shows a power-law behaviour with a somewhat larger power of -3.6, but clearly the behaviour of conventional systems qualitatively differs from the perovskite-based superconductors.

*Ru inclusions in SRO.* Strontium ruthenate single crystals often contain inclusions of ruthenium metal due to the growth conditions and necessary excess of Ru. It is known that $T_c$ is locally enhanced around the inclusions[39] (within the SRO, not the Ru metal), most likely due to the induced local stress[24]. Crystals with the highest achievable $T_c$ ~ 1.5 K can still contain a non-negligible concentration of Ru inclusions, i.e., their number is unrelated to the level of point defects. Due to the strong local enhancement of $T_c$, it is possible that the material around inclusions contributes to the nonlinear susceptibility far above the bulk $T_c$. As a result of the high sensitivity of our technique, it can detect inclusion-related signals at the level of $10^{-3}$ Ru volume fraction or lower. However, it seems unlikely that the inclusions would qualitatively change the nonlinear response in the entire range from $T_c$ to ~ 2 K, especially close to $T_c$ (the transition is sharp and well defined in the linear susceptibility; see Fig. S2 inset). Conceivably, the high-temperature tail of the response is influenced by the inclusions, which might explain the apparent change in slope of the exponential dependence (Fig. S2). Furthermore, we compare the nonlinear susceptibility with specific heat measurements[40], where an exponential tail is observed as well (Fig. S2). The extracted characteristic temperature $\Xi$ is about 50% smaller than the equivalent slope of the nonlinear susceptibility close to $T_c$, and approximately two times smaller than the high-temperature slope. Importantly, the temperature dependence is qualitatively the same.

In an effort to further investigate the possible influence of inclusions, we have performed nonlinear response measurements on a sample from a different growth, and with a lower $T_c$ of 1.33 K (Fig. S3). The slopes of the exponential tails are virtually the same for the two samples, despite the difference in $T_c$ and crystal growth conditions, which suggests that the observed behaviour is at best weakly influenced by Ru inclusions. We further compare the results to measurements on the same $T_c$ = 1.51 K after additional polishing (in an attempt to remove some inclusion-containing material). Again, no significant difference is observed.

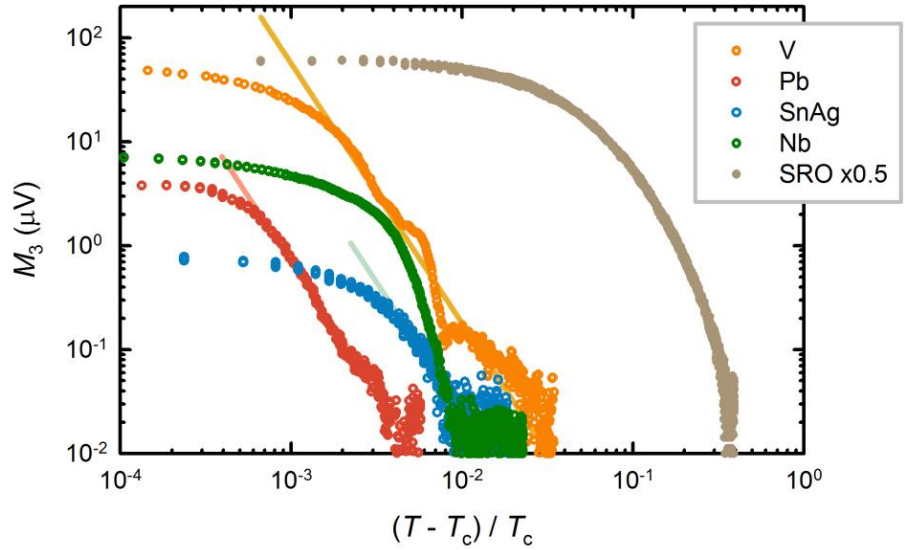

**Fig. S1 | Nonlinear magnetization of four conventional superconductors**. The nonlinear response of the conventional superconductors vanadium, lead, tin-silver alloy and niobium single crystal is compared to that for strontium ruthenate on a relative, reduced temperature scale. The kink in the vanadium data is possibly a spurious signal due to solder superconductivity, or the result of slight macroscopic sample inhomogeneity. V, Pb and SnAg follow the mean-field Ginzburg-Landau prediction[11] of a power-law tail with exponent -5/2, whereas the exponent for niobium is slightly larger at -3.6. The SRO data, in contrast, do not show power-law behaviour at any reduced temperature, and extend to much higher reduced temperatures.

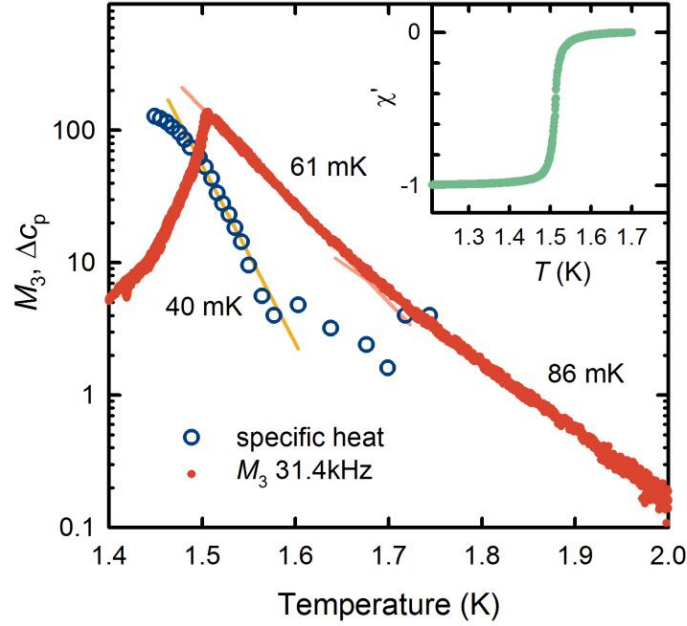

**Fig. S2 | SRO linear susceptibility and specific heat.** Nonlinear response (full circles) compared to electronic specific heat data (empty circles, from ref. 40) with the constant normal-state specific heat contribution subtracted. Exponential dependences are found in both observables, with the slope of the specific heat data (~ 40 mK) somewhat steeper. The inset shows the linear susceptibility of the SRO sample and demonstrates a sharp superconducting transition.

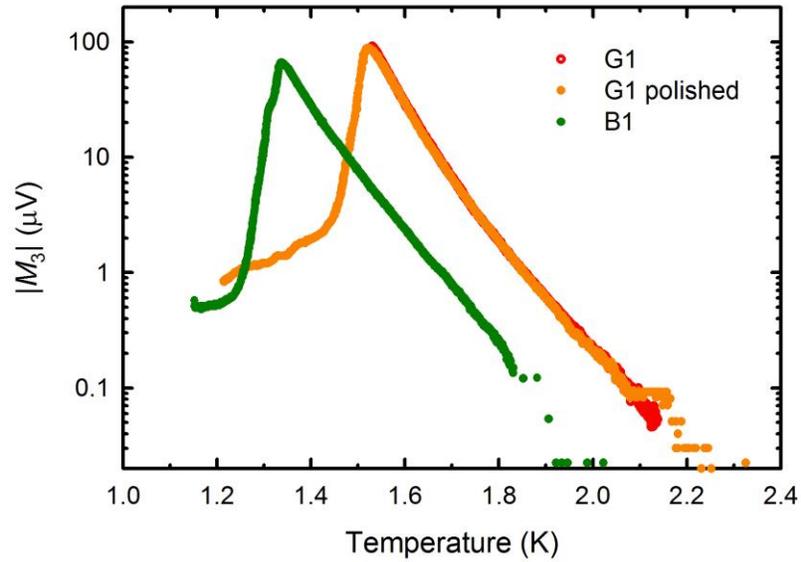

**Fig. S3 | Comparison of different $Sr_2RuO_4$ samples.** Measurements of third-order response for two SRO samples with $T_c$ of 1.51 K (sample G1) and 1.33 K (sample B1) demonstrate closely similar behaviour above their respective $T_c$. A comparison of measurements of the G1 crystal before and after additional polishing is shown as well. The polishing was done with the intent to remove some inclusion-containing material, yet no difference in the response is discerned.